\documentstyle[11pt,a4wide]{article}

\begin{document}
\vspace{.5cm}
\centerline{\bf\large{\bf{Hamiltonian Structures of Multi-component Constrained
KP Hierarchy }}}
\vspace{1in}
\centerline{\bf\large Q.\ P. \ Liu}
\smallskip
\smallskip
\centerline{\bf\large CCAST(World Laboratory),}
\centerline{\bf\large P.O.Box 8730, Beijing 100080, China,}
\smallskip
\smallskip
\centerline{\bf\large and}
\smallskip
\smallskip
\centerline{\bf\large Department of Mathematics, }
\centerline{\bf\large Beijing Graduate School,}
\centerline{\bf\large China University of Mining and Technology, }
\centerline{\bf\large Beijing 100083, China\footnote{Mailing Address}}
\newcommand{\be}{\begin{equation}}
\newcommand{\ee}{\end{equation}}
\newcommand{\ba}{\begin{array}}
\newcommand{\ea}{\end{array}}
\renewcommand{\theequation}{\thesection, \arabic{equation}}

\def\p{\partial}
\def\alf{\alpha}
\def\bi{\beta}
\def\es{\epsilon}
\def\la{\lambda}
\vspace{0.5in}\begin{center}
\begin{minipage}{5in}
{\bf ABSTRACT}\hspace{.2in}We consider the Hamiltonian theory for the
multi-component KP hierarchy. We show that the second Hamiltonian structures
constructed by Sidorenko and Strampp[J. Math. Phys. {\bf 34}, 1429(1993)] are
not Hamiltonian. A candidate for the second Hamiltonian Structures is proposed
and is proved to lead to hereditary operators.\par
\end{minipage}
\end{center}
\vspace{.5in}
\vfill\eject
\par
\section{INTRODUCTION}
\setcounter{equation}{0}
During last few years, the constrained KP hierarchy is studied
intensively$^{1-3,5}$. This hierarchy is the result of generalizing Cao's$^6$
nonlinearization to the 2+1 dimensional case. Very interestingly, as the famous
Gelfand-Dikii hierarchy$^7$, the constrained KP hierarchy is both
mathematically and physically important. On the one hand, it contains
physically applicable models, such as Yajima-Oikawa$^8$ model and Melnikov$^9$
model. On the other hand, the constrained KP hierarchy is Bi-Hamiltonian$^5$,
has Darboux transformation$^{10}$, can be modified$^{11}$ and is relevant to
the theory of the W algebra$^{12}$. Very recently, the constrained KP is shown
to be just a special case of a more general restriction of the KP
hierarchy$^{13}$.\par
Sidorenko and Strampp$^4$ introduced multi-component KP hierarchy, which is a
straightforward generalization of the scalar case. This is the hierarchy
associated with the following Lax operator
\be
L_n=\p^n+u_{n-2}\p^{n-2}+...+u_0+\sum^{m}_{i=1}q_i\p^{-1} r_i,
\ee
the corresponding flows may be constructed by means of Fractional Power
Method$^7$. For $n=1$, one has multi-component AKNS, which includes the
important coupled nonlinear Schr\"{o}dinger equation$^{14}$ as a special case.
For the cases $n=2$ and $n=3$, one has the multi-component Yajima-Oikawa
hierarchy and Melnikov hierarchy respectively. Sidorenko and Strampp$^4$
further constructed recursion operators for the cases $n=2$ and $n=3$ by means
of variational calculus developed in the Ref.15. They claimed that their
recursion operators have implectic-symplectic factorizations in the sense of
Fuchssteiner and Fokas$^{17}$. That is to say, they claimed that the
Bi-Hamiltonian structures are found for the multi-component Yajima-Oikawa
hierarchy and Melnikov hierarchy. Unfortunately, they did not prove their
statement either directly or indirectly. The partial reason is that their
candidates for the Hamiltonian structures are complicated nonlocal matrix
operators and direct proof would be too tedious to do by ha

The aim of the paper is to show that Sidorenko and Strampp's claim is not
correct. We will prove that their second Hamiltonian structures are not
qualified as Hamiltonian at all. Furthermore, we will present alternative
candidates for the second Hamiltonian structures and prove that it leads to
hereditary operators. For simplicity, we concentrate on the simplest and non
trivial case: two-component case. The generalization to the multi-component
case will be commented in the due course.

The paper is arranged as follows. The next section is on the two-component AKNS
systems. Sidorenko-Strampp type operator is presented and shown to be not
hereditary. A candidate for the second Hamiltonian operators is constructed.
Also, we present a hereditary operator for this hierarchy. We do the same thing
in the section three for the two-component Yajima-Oikawa hierarchy. Section
four contains some comments on generalizations of the results of section two
and section three and gives some outlines for the further study.

\section{TWO-COMPONENT AKNS HIERARCHY}
\setcounter{equation}{0}
We consider the two-component AKNS system next. This hierarchy is known for
long time. In fact, the so important coupled nonlinear Schr\"{o}dinger
equation$^{14}$ is a reduction of it. However, the explicit form of recursion
operator is not written down to the best of my knowledge although it might be
known to the specialists. For the motivation of the next section, we present it
here.

We first give Sidorenko-Strampp type operator and show it is not hereditary.

The Lax operator is
\be
L_1=\p-q_1\p^{-1}r_1-q_2\p^{-1}r_2,
\ee
the flows are
\be
L_{1_{t_{k}}}=[((L_1)^k)_+, L_1],
\ee
where subscript + means the projection to the diffenertial part.

Following the idea of Sidorenko and Strampp$^4$, we easily see that the
systems(2.2) have the following presentation

\be
{\bf q}_{t_{k}}={\bf B}_{{ss}_{0}} \frac{\delta H_{k+1}}{\delta{\bf q}}={\bf
B}_{{ss}_{1}} \frac{\delta H_{k}}{\delta{\bf q}},
\ee
where ${\bf q}=(q_1,q_2,r_1,r_2)^T$ and
\be
{\bf B}_{{ss}_{0}}= \left[\ba{cccc}  0&0&1&0\\ 0&0&0&1\\ -1&0&0&0\\ 0&-1&0&0\ea
\right],
\ee
\be
{\bf B}_{{ss}_{1}}=\left[\ba{cccc}2q_1\p^{-1}q_1&
2q_1\p^{-1}q_2&\p-2q_1\p^{-1}r_1
&-2q_1\p^{-1}r_2\\
2q_2\p^{-1}q_1&2q_2\p^{-1}q_2&-2q_2\p^{-1}r_1&\p-2q_2\p^{-1}r_2\\
\p-2r_1\p^{-1}q_1&-2r_1\p^{-1}q_2&2r_1\p^{-1}r_1&2r_1\p^{-1}r_2\\
-2r_2\p^{-1}q_1&\p- 2r_2\p^{-1}q_2&2r_2\p^{-1}r_1&2r_2\p^{-1}r_2\ea \right],
\ee
and Hamiltonian functionals $H_k$ may be calculated from the formula:
\be
H_k=\frac{1}{k}(Res(L_1)^k).
\ee
The Hamiltonian nature of ${\bf B}_{{ss}_{0}}$ is self-evident. However, it is
not clear that if ${\bf B}_{{ss}_{1}}$ is or not a Hamiltonian operator
although it is skew symmetric. Next, we prove that ${\bf B}_{{ss}_{1}}$ is not
Hamiltonian. To show this, let us first simplify ${\bf B}_{{ss}_{1}}$ via
coordinate transformations. Motivated by the situation in the scalar
case$^{12}$, we introduce the following coodinates
\be
S_1=q_1r_1,\quad S_2=q_2r_2,\quad T_1=-\frac{r_{1x}}{r_1},\quad
T_2=-\frac{r_{2x}}{r_2},
\ee
then, it is ready to see that ${\bf B}_{{ss}_{0}}$ and ${\bf B}_{{ss}_{1}}$ are
transformed to
\be
\hat{{\bf B}}_{{ss}_{0}}= \left[\ba{cccc}  0&0&\p&0\\ 0&0&0&\p\\ \p&0&0&0\\
0&\p&0&0\ea \right],
\ee
\be
\hat{{\bf B}}_{{ss}_{1}}=\left[\ba{cccc}S_1\p+\p S_1&0&\p^2+T_1\p&0\\
0&S_2\p+\p S_2&0&\p^2+T_2\p\\ -\p^2+\p T_1&0&-2\p&-2\p\\ 0& -\p^2+\p
T_2&-2\p&-2\p\ea \right],
\ee
thus, the transformation(2.7) localizes ${\bf B}_{{ss}_{0}}$ and ${\bf
B}_{{ss}_{1}}$. With  (2.8-9) in hand, we have a recursion operator

\be
{\bf R}_{ss}=\left[\ba{cccc} \p+T_1&0&2S_1+S_{1x}\p^{-1}&0\\
0&\p+T_2&0&2S_2+S_{2x}\p^{-1}\\ -2&-2&-\p+\p T_1\p^{-1}&0\\ -2&-2&0&-\p+\p
T_2\p^{-1}\ea \right].
\ee
Now, if ${\bf B}_{{ss}_{1}}$ would be a Hamiltonian operator, one would have a
hereditary operator in the sense of Fuchssteiner$^{16}$. That means the
following identity must be hold
\be
{\bf R}_{ss}^{'}[{\bf R}_{ss}(f)]g-{\bf R}_{ss}^{'}[{\bf R}_{ss}(g)]f={\bf
R}_{ss}({\bf R}_{ss}^{'}[f]g-{\bf R}_{ss}^{'}[g]f),
\ee
for arbitrary vector function f and g. Where $'$ denotes Gateaux derivative.

However, a long calculation shows that it is not the case here. In fact,
letting $f=(f_1,f_2,f_3,f_4)^T$ and $g=(g_1,g_2,g_3,g_4)^T$, we have
\be
{\bf R}_{ss}^{'}[{\bf R}_{ss}(f)]g-{\bf R}_{ss}^{'}[{\bf R}_{ss}(g)]f-{\bf
R}_{ss}({\bf R}_{ss}^{'}[f]g-{\bf R}_{ss}^{'}[g]f)=2(-f_2g_1+g_2f_1),
\ee
which is not identical zero. This means that ${\bf R}_{ss}$ is not hereditary.
So we conclude that ${\bf B}_{{ss}_{1}}$ is not Hamiltonian.

In the remain part of the section, we construct a hereditary operator for the
hierarchy(2.2). To do this, let us rewrite the corresponding spectral problem
as matrix form
\be
\left[\ba{ccc}\phi\\ \phi_1\\ \phi_2\ea
\right]_x=\left[\ba{ccc}\la&q_1&q_2\\r_1&0&0\\ r_2&0&0\ea
\right]\left[\ba{ccc}\phi\\ \phi_1\\ \phi_2\ea \right] \equiv U\Phi,
\ee
as usual, we adjoin (2.13) with time evolution of wave function $\Phi$:
$\Phi_t=V\Phi$. Then, calculating zero-curvature equation $U_t-V_x+[U,
V]=0$ leads us to
\be
{\bf q}_{t_{k}}={\bf B}_0\frac{\delta H_{k+1}}{\delta{\bf q}}={\bf
B}_1\frac{\delta H_{k}}{\delta{\bf q}},
\ee
where
\be
{\bf B}_0={\bf B}_{{ss}_{0}},
\ee
\be
{\bf
B}_1=\left[\ba{cccc}2q_1\p^{-1}q_1&q_1\p^{-1}q_2+q_2\p^{-1}q_1&R_1&-q_1\p^{-1}r_2\\
q_1\p^{-1}q_2+q_2\p^{-1}q_1&2q_2\p^{-1}q_2&-q_2\p^{-1}r_1&R_2\\
-(R_1)^*&-r_1\p^{-1}q_2&2r_1\p^{-1}r_1&r_1\p^{-1}r_2+r_2\p^{-1}r_1\\
-r_2\p^{-1}q_1&-(R_2)^*&r_1\p^{-1}r_2+r_2\p^{-1}r_1&2r_2\p^{-1}r_2\ea \right].
\ee
\be
R_1=\p-2q_1\p^{-1}r_1-q_2\p^{-1}r_2,\quad R_2=\p- q_1\p^{-1}r_1-2q_2\p^{-1}r_2,
\ee
Hamiltonians $H_n$ may be calculated as before. To say that ${\bf B}_1$ is a
Hamiltonian operator requires rather tedious calculation. One may suppose it is
localizable, but the transformation(2.7) certainly does not do this job and I
am not able to find such transformation at present. Here we are not going to
prove the Hamiltonian nature of ${\bf B}_1$ directly or indirectly, although we
believe it is the case. Instead, we form a recursion operator
\smallskip
\be
{\bf
R}=\left[\ba{cccc}R_1&-q_1\p^{-1}r_2&-2q_1\p^{-1}q_1&-q_1\p^{-1}q_2-q_2\p^{-1}q_1\\
 -q_2\p^{-1}r_1&R_2&-q_1\p^{-1}q_2-q_2\p^{-1}q_1&-2q_2\p^{-1}q_2\\
 2r_1\p^{-1}r_1&r_1\p^{-1}r_2+r_2\p^{-1}r_1&(R_1)^*&r_1\p^{-1}q_2\\
 r_1\p^{-1}r_2+r_2\p^{-1}r_1&2r_2\p^{-1}r_2&r_2\p^{-1}q_1&(R_2)^*\ea \right],
\ee
where $R_1$ and $R_2$ are defined by (2.17).
\smallskip

Then, straightforward but cumbersome calculation shows that ${\bf R}$ is indeed
hereditary. This also supports our conjecture: ${\bf B}_1$ is Hamiltonian.
\section{COUPLED YAJIMA-OIKAWA HIERARCHY}
\setcounter{equation}{0}
As promised in the Introduction, we prove that Sidorenko-Strampp's operator(see
Ref. 4) is not Hamiltonian. After this, we give a candidate for the second
Hamiltonian operator.

The Lax operator in this case is
\be
L_2=\p^2-u-q_1\p^{-1}r_1-q_2\p^{-1}r_2,
\ee
two Hamiltonian operators given by Sidorenko and Strampp$^4$ for the hierarchy
associated with (3.1) are
\be
{{\bf B}}_{{ss}_{0}}= \left[\ba{ccccc} -2\p&0&0&0&0\\  0&0&0&1&0\\ 0&0&0&0&1\\
0&-1&0&0&0\\ 0&0&-1&0&0\ea \right],
\ee

\be
{{\bf B}}_{{ss}_{1}}=
\left[\ba{ccccc}-\frac{1}{2}\p^3+u\p+\p u&q_1\p+\frac{1}{2}\p
q_1&q_2\p+\frac{1}{2}\p q_2&r_1\p+\frac{1}{2}\p r_1& r_2\p+\frac{1}{2}\p r_2\\
\p
q_1+\frac{1}{2}q_1\p&\frac{3}{2}q_1\p^{-1}q_1&\frac{3}{2}q_1\p^{-1}q_2&J(q_1,r_1)&-\frac{3}{2}q_1\p^{-1}r_2\\
 \p
q_2+\frac{1}{2}q_2\p&\frac{3}{2}q_2\p^{-1}q_1&\frac{3}{2}q_2\p^{-1}q_2&-\frac{3}{2}q_2\p^{-1}r_1&J(q_2,r_2)\\
\p
r_1+\frac{1}{2}r_1\p&-(J(q_1,r_1)^*&-\frac{3}{2}r_1\p^{-1}q_2&\frac{3}{2}r_1\p^{-1}r_1&\frac{3}{2}r_1\p^{-1}r_2\\
\p
r_2+\frac{1}{2}r_2\p&-\frac{3}{2}r_2\p^{-1}q_1&-(J(q_2,r_2)^*&\frac{3}{2}r_2\p^{-1}r_1&\frac{3}{2}r_2\p^{-1}r_2\ea \right],
\ee
with $J(q,r)\equiv \p^2-u-\frac{3}{2}q\p^{-1}r$
\smallskip

We notice that ${\bf B}_{{ss}_{1}}$ is exactly the one given in the Ref.4 apart
from a scaling. As before, we introduce new coordinates
\be
u=u,\quad S_1=q_1r_1,\quad S_2=q_2r_2,\quad T_1=-\frac{r_{1_{x}}}{r_1},\quad
T_2=-\frac{r_{2_{x}}}{r_2},
\ee
then, ${\bf B}_{ss_{i}}$ take the following forms
\be
\hat{{\bf B}}_{{ss}_{0}}= \left[\ba{ccccc} -2\p&0&0&0&0\\  0&0&0&\p&0\\
0&0&0&0&\p\\ 0&\p&0&0&0\\ 0&0&\p&0&0\ea \right],
\ee
\be
\hat{{\bf B}}_{{ss}_{1}}=
\left[\ba{ccccc}-\frac{1}{2}\p^3+u\p+\p
u&3S_1\p+2S_{1x}&3S_2\p+2S_{2x}&T_1\p+\frac{3}{2}\p^2&T_2\p+\frac{3}{2}\p^2\\
3S_1\p+S_{1x}&J_1 (S_1,T_1)&0&J_2 (T_1)&0\\
3S_2\p+S_{2x}&0&J_1 (S_2,T_2)&0&J_2 (T_2)\\
 \p T_1-\frac{3}{2}\p^2&J_3 (T_1)&0&-\frac{3}{2}\p&-\frac{3}{2}\p\\
\p T_1-\frac{3}{2}\p^2&0&J_3 (T_2)&-\frac{3}{2}\p&-\frac{3}{2}\p\ea \right].
\ee
with $J_1 (S,T) \equiv (S_x+2ST)\p+\p(S_x+2ST)$, $J_2 (T)\equiv
(T^2+T_{1x}+2T_1\p+\p^2)\p-u\p$ and $J_3 (T) \equiv -(J_2(T))^*$.
\smallskip

Exactly as above, we found that the operator ${\bf R_{ss}}={\bf
B}_{{ss}_{1}}({\bf B}_{{ss}_{0}})^{-1}$ is not hereditary. Therefore, ${\bf
B}_{ss_{1}}$ is not Hamiltonian.

As in the AKNS case of last section, we may use the zero curvature equation and
derive the following representation of the hierarchy
\be
{\bf u}_{t_{k}}={\bf B}_0\frac{\delta H_{k+1}}{\delta{\bf u}}={\bf
B}_1\frac{\delta H_k}{\delta{\bf u}},
\ee
where ${\bf u}=(u,q_1,q_2,r_1,r_2)^T$ and ${\bf B}_0={\bf B}_{ss_{0}}$

\be
{\bf B}_1=
\left[\ba{ccccc}-\frac{1}{2}\p^3+u\p+\p u&q_1\p+\frac{1}{2}\p
q_1&q_2\p+\frac{1}{2}\p q_2&r_1\p+\frac{1}{2}\p r_1& r_2\p+\frac{1}{2}\p r_2\\
\p q_1+\frac{1}{2}q_1\p&\frac{3}{2}q_1\p^{-1}q_1&I(q_1,q_2)
&I_1&-\frac{1}{2}q_1\p^{-1}r_2\\
 \p q_2+\frac{1}{2}q_2\p&
I(q_2,q_1)&\frac{3}{2}q_2\p^{-1}q_2&-\frac{1}{2}q_2\p^{-1}r_1&I_2\\
\p
r_1+\frac{1}{2}r_1\p&-I_{1}^{*}&-\frac{1}{2}r_1\p^{-1}q_2&\frac{3}{2}r_1\p^{-1}r_1&I(r_1,r_2)\\
\p
r_2+\frac{1}{2}r_2\p&-\frac{1}{2}r_2\p^{-1}q_1&-I_{2}^{*}&I(r_2,r_1)&\frac{3}{2}r_2\p^{-1}r_2\ea \right].
\ee
where $I(v_1,v_2)\equiv \frac{1}{2}v_1\p^{-1}v_2 +v_2\p^{-1}v_1$, $I_1\equiv
\p^2-u-\frac{3}{2}q_1\p^{-1}r_1-q_2\p^{-1}r_2$ and $I_2 \equiv
\p^2-u-\frac{3}{2}q_2\p^{-1}r_2-q_1\p^{-1}r_1$
\smallskip

Direct verification of the Hamiltonian nature of ${\bf B}_1$ is too tedious to
perform directly. Instead, we performed calculation on verification of the
hereditary of the corresponding recursion operator. It should be noticed that
the verification of hereditary is simpler. With a hereditary operator in hand,
standard theory$^{16}$ allows us to construct commuting flows.

\section{CONCLUSIONS}
We considered the Hamiltonian theory for the multi-component constrained KP
hierarchy. It is proved, in the simplest non trivial case, that the Hamiltonian
operators calculated by Sidorenko and Strampp$^4$ are by no means Hamiltonian.
Alternative Hamiltonian structures are proposed and they are shown to lead to
hereditary operators.

The above results may be generalized along two directions:  generic
multi-component case and higher order constrained KP case. While both
generalizations are straightforward, the calculations will be extremely
involved. Here, we just comment that the the results presented in the Appendix
of the Ref.4 are not correct.

There are several points which deserves further consideration. We list them
here: (1). Hamiltonian nature of the second structures(2.16) and(3.8). As
pointed above, a direct verification is too cumbersome to do by bare hand. This
may be completed by a symbolic program, such as Maple or Mathematica. Another
way to do this is to use a Miura map, which should simplify the second
structure considerablely if it exists. Finding a Miura map is interesting in
its own right; (2). Modifications of the hierarchies presented here. In the
scalar case, this problem is solved in the Ref.11. The generalization to the
multi-component case is important; (3).Associated classical W algebras. Once
again, this problem in the scalar case is solved$^{12}$. Here we just remark
that the structure(3.6) is closely related to so-called bosonic analogy of the
Knizhnik-Bershadsky superconformal algebra$^{18}$. Some of these problems are
under investigation.

\vspace{10pt}
{\bf{ACKNOWLEDGEMENT}}\par
This work is supported by Natural National Science Foundation of China.\par
\vspace{.1in}
\bigskip

\large
REFENERCES
\smallskip
\par
\small
\begin{flushleft}
$^1$ Y. Cheng, J. Math. Phys. {\bf 33}, 3774(1992).\par
$^2$ B.G. Konopelchenko and W. Strampp, Inverse Problems {\bf 7}, L17(1991);
B.G. Konopelchenko, J. Sidorenko and W. Strampp, Phys. Lett. A {\bf 157},
17(1991); J. Sidorenko and W. Strampp, Inverse Problems {\bf 7}, L37(1991).\par
$^3$ Y. Cheng and Y. Li, Phys. Lett. A {\bf 157}, 22(1991).

$^4$ J. Sidorenko and W. Strampp, J. Math. Phys. {\bf 34}, 1429(1993).\par
$^5$ W. Oevel and W. Strampp, Commun. Math. Phys. {\bf 157}, 51(1993).

$^6$ C.W. Cao, Henan Science {\bf 2},  2(1987); Science in China(Scientia
Sinica)  {\bf 33}, 528(1990).\par
$^7$ L.A. Dickey, {\em Soliton equations and Hamiltonian systems}, World
Scientific, Singapore(1991).

$^8$ N. Yajima and M. Oikawa, Progr. Theor. Phys. {\bf 56}, 1719(1976).\par
$^9$ V.K. Melnikov, Phys. Lett.A {\bf 118}, 22(1986); Lett. Math. Phys. {\bf
7}, 129(1983); Commun. Math. Phys. {\bf 112}, 639(1987); Commun. Math. Phys.
{\bf 120}, 481(1989).\par
$^{10}$ W. Oevel, Physica A {\bf 195},, 533(1993).

$^{11}$ Q.P. Liu, Phys. Lett.A  {\bf 187} 373(1994).  \par
$^{12}$ Q.P. Liu and C.S. Xiong, Phys. Lett.B {\bf 327}, 257(1994).\par
$^{13}$ M. Borona, Q.P. Liu and C.S. Xiong, Bonn-Th-9417, SISSA-ISAS-118/94,
AS-ITP-94-43, hep-th/9408035.

$^{14}$ S. Manakov, Sov. Phys. JEPT  {\bf 38}, 248(1974).\par
$^{15}$ B.A. Kupershmidt, {\em Discrete Lax equations and
differential-difference calculus}, Revue Asterisque, Vol. 123, Paris(1985).

$^{16}$ B. Fuchssteiner, Nonlinear Anal. Theory Method Appl. {\bf 3},
849(1979).

$^{17}$ B. Fuchssteiner and A.S. Fokas, Physica D {\bf 4}, 47(1981).

$^{18}$ J. Fuchs, Phys. Lett.B {\bf 262} 249(1991);  L.J. Romans, Nucl. Phys. B
{\bf 357}, 549(1991).
\end{flushleft}
\par
\end{document}